\documentclass{aastex631}

\begin{document}

\title{BSN: First Light Curve Study of the Low Mass Contact Binary V0610 Vir}

\author[0000-0001-5768-0340]{Ailar Alizadehsabegh}
\affiliation{Department of Photonics, University of Tabriz, Tabriz, Iran}

\author{František Lomoz}
\affiliation{Variable Star and Exoplanet Section, Czech Astronomical Society, Prague, Czech Republic}

\author[0000-0002-0196-9732]{Atila Poro}
\affiliation{Astronomy Department of the Raderon AI Lab., BC., Burnaby, Canada}

\author{Ata Narimani}
\affiliation{Independent Researcher, Tabriz, Iran}

\begin{abstract}
Photometric data were used to perform the first light curve analysis of the V0610 Vir binary system. Observations and analysis were done in the form of the Binary Systems of South and North (BSN) Project. We extracted the minima from our observations and compiled the literature, which was few. Therefore, we performed computations using the reference ephemeris and presented a new ephemeris and O-C diagram with a linear fit. Light curve analysis was performed using the PHOEBE Python code and the Markov chain Monte Carlo (MCMC) approach. The assumption of a hot starspot was required due to the asymmetry in the light curve's maxima. The analysis shows that V0610 Vir is a contact binary system with a fillout factor of 0.085, a mass ratio of 0.998, and an inclination of $70.65^{\circ}$. The absolute parameters of the system were estimated based on the Gaia DR3 parallax method. The results show that the system is a low-mass contact binary with a total mass lower than $0.8(M_\odot)$. The location of the stars was shown in the $M-L$ and $M-R$ diagrams.
\end{abstract}


\vspace{1.5cm}
\section{Introduction}
The W Ursae Majoris (W UMa) binaries include two stars that are typically F, G, or K spectral type stars whose Roche lobes have been fielded, and they share a similar envelope (\citealt{2016RAA....16..135K}[1], \citealt{2016PASP..128l4201Z}[2], \citealt{2024PASP..136b4201P}[3]). The orbital period of W UMa-type systems is less than one day. Also, the light curves of these binary stars show two equal or almost equal minima, demonstrating that their effective surface temperatures are close to each other (\citealt{2018PASP..130g4201L}[4]). Further investigation of contact systems is important since it can reveal many details about the evolution of stars.

V0610 Vir listed in the \cite{2015IBVS.6151....1K}[5] study. According to the AAVSO International Variable Star Index (VSX\footnote{\url{https://www.aavso.org/vsx/}}) and ASAS-SN\footnote{\url{http://asas-sn.osu.edu/variables}} variable stars’ catalogs, V0610 Vir is a W UMa-type binary system with an orbital period of $0.3398754^{days}$ and $0.3398768^{days}$, respectively. The coordinates of this system in the Gaia DR3\footnote{\url{https://gea.esac.esa.int/archive/}} database are R.A.: $176.7745788^{\circ}$ and Dec.: $1.2447709^{\circ}$ (J2000).

The maximum apparent magnitude $V_{max}$ of the system was reported as $13.31^{mag}$ in the ASAS3 catalog, $13.15^{mag}$ in the GCVS\footnote{General Catalogue of Variable Stars} catalog, $13.31^{mag}$ in the AAVSO\footnote{American Association of Variable Star Observers (AAVSO), International Variable Star Index (VSX)}, and $13.15^{mag}$ in the \cite{2015IBVS.6151....1K}[5] study.

This work is a continuation of the BSN\footnote{\url{https://bsnp.info/}} project on eclipsing binary systems. The paper's structure is as follows: Section 2 explains observation and data redaction; Section 3 is about extracting minima and obtaining new ephemeris; and Section 4 is related to light curve analysis. The technique used to estimate the absolute parameters is described in Section 5, and the conclusion is in Section 6.

\vspace{1.5cm}
\section{Observation and Data Reduction}
The observations in the photometric system of V0610 Vir were carried out one night in March 2020 by a Schmidt-Newton 254mm/1016mm telescope with the G2-8300 CCD camera at a private observatory in the Czech Republic (49.65 N, 14.41 E). During observations, the CCD average temperature was $-20^{\circ}C$.

A $V$-band filter was used, and a total of 279 images were obtained. Each image has an exposure time of 90 seconds. Images were processed using MaxIm DL software, which included dark, bias, and flat-field for basic data reduction.

\begin{figure*}
\begin{center}
\includegraphics[scale=0.45]{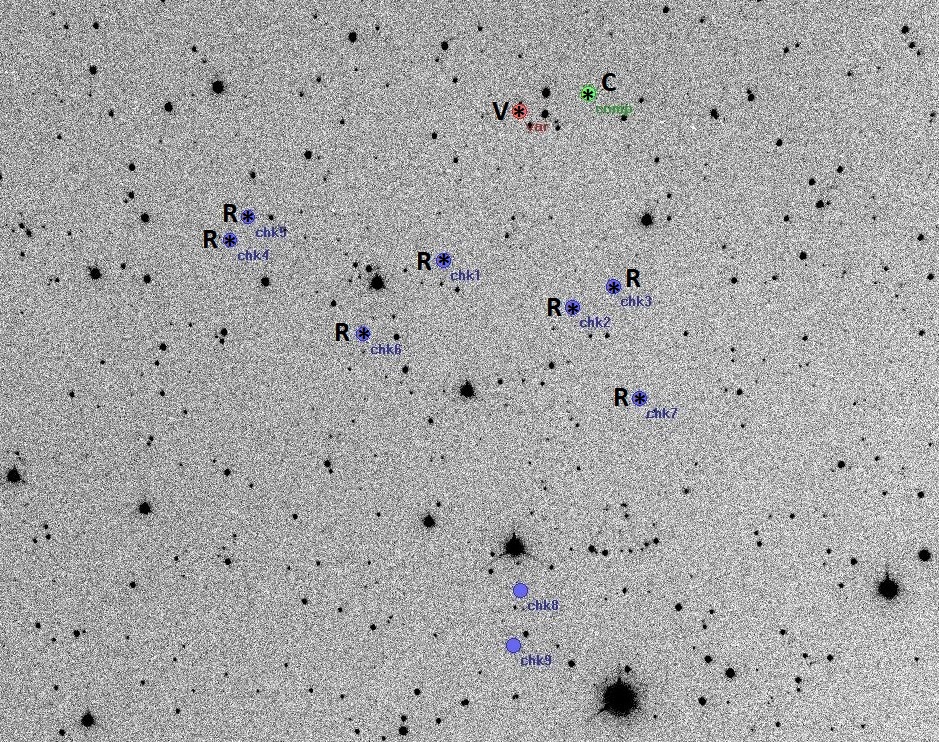}
    \caption{Field-of-view of the V0610 Vir binary system (V), comparison (C), and check stars (R). The field-of-view of the picture is $35\times25$ arcminutes.}
\label{Fig1}
\end{center}
\end{figure*}

Figure \ref{Fig1} displays the comparison and check stars that were selected that were close to the target and had a suitable apparent magnitude in comparison to V0610 Vir.
So, we considered a comparison star named UCAC4 459-049136 (11 46 55.388, +01 44 51.534) with an apparent magnitude of $V=14.32$ and nine check stars including UCAC4 457-049442 (11 47 17.678, +01 20 02.553) with a $V=13.20$ magnitude, UCAC4 458-049431 (11 46 59.259, +01 21 57.295) with a $V=13.72$ magnitude, UCAC4 457-049428 (11 46 52.883, +01 21 12.000) with a $V=13.29$ magnitude, UCAC4 457-049456 (11 47 48.705, +01 19 04.569) with a $V=13.90$ magnitude, UCAC4 457-049455 (11 47 46.001, +01 18 14.696) with a $V=14.51$ magnitude, UCAC4 457-049448 (11 47 29.787, +01 22 38.372) with a $V=14.13$ magnitude, UCAC4 458-051054 (11 46 49.701, +01 25 18.116) with a $V=14.17$ magnitude, UCAC4 458-051065 (11 47 07.913, +01 32 12.092) with a $V=13.65$ magnitude and UCAC4 458-051064 (11 47 07.423, +01 33 47.439) with a $V=15.37$ magnitude. The coordinates and apparent magnitudes of all the comparison and check stars were gathered from the ASAS-SN catalog.

Finally, we used the AstroImageJ program to normalize the flux of all the data (\citealt{2017AJ....153...77C}[6]).

\vspace{1.5cm}
\section{New Ephemeris}
\begin{figure*}
\begin{center}
\includegraphics[scale=0.60]{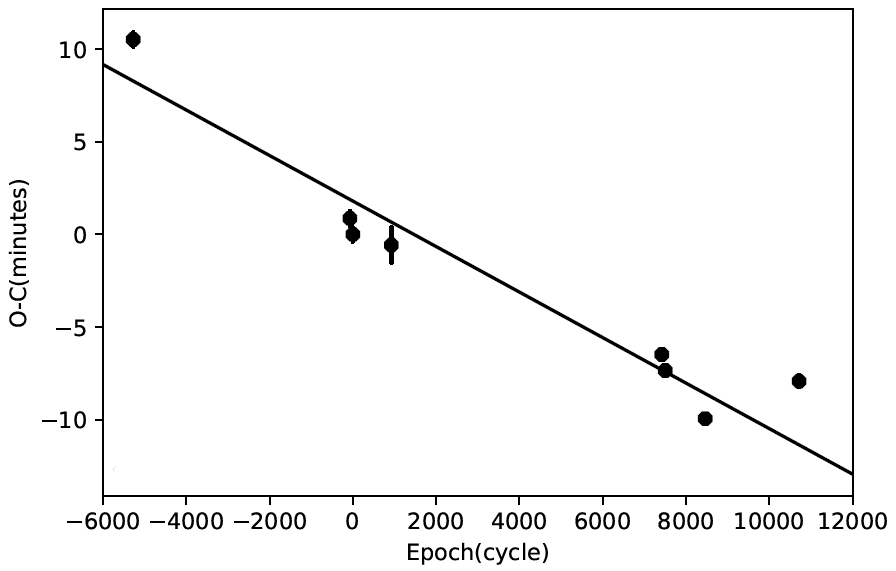}
    \caption{O-C diagram with a linear fit for the V0610 Vir system.}
\label{Fig2}
\end{center}
\end{figure*}

In binary star systems, the O-C diagram is an important tool for finding new ephemeris. The O-C value represents the difference between the predicted and observed times of an eclipse in a binary system. The O-C diagram allows us to visualize the changes in the timing of eclipses over a longer period of time. By analyzing the trends of the O-C diagram, we can study various phenomena, including orbital period variations, eclipse timing variations, and even the presence of additional companions in the system.
For this purpose, we extracted one primary and one secondary from our observations and collected them with ten other minima from the literature (Table \ref{tab1}). Also, we converted all of the times of minima to the Barycentric Julian Date in Barycentric Dynamical Time ($BJD_{TDB}$)\footnote{\url{https://astroutils.astronomy.osu.edu/time/}}.
\\
\\
To compute Epoch and O-C, we used a reference ephemeris with a time of minima of 2455291.81179(30) from \cite{2010IBVS.5945....1D}[7] and an orbital period of $0.3398768^{days}$ that we obtained from the ASAS-SN catalog. Therefore, according to the O-C diagram and considering that the number of observations for this system is limited and few minima are available for it, only a least-squares linear fit can be considered (Figure \ref{Fig2}). Based on this information, the new ephemeris can be calculated as follows:

\begin{equation}\label{eq1}
Min.I (BJD_{TDB})=2455291.81304(13)+0.339875947(17)\times E
\end{equation}

where $Min.I$ is related to the primary minimum, and $E$ is the cycle.

\begin{table*}
\caption{Available times of minima for V0610 Vir.}
\centering
\begin{center}
\footnotesize
\begin{tabular}{c c c c c c}
 \hline
 \hline
Min.($BJD_{TDB}$) & Error & Method & Epoch & O-C(day) & Reference\\
\hline
2453499.64872 & 0.00030 & $V$ & -5273 & 0.0073 & \cite{2010IBVS.5945....1D}[7]\\
2455267.68109 & 0.00030 & $V$ & -71 & 0.0006 & \cite{2010IBVS.5945....1D}[7]\\
2455291.81179 & 0.00030 & $V$ & 0 & 0 & \cite{2010IBVS.5945....1D}[7]\\
2455605.85759 & 0.00070	& $V$ & 924 & -0.0004	& \cite{2011IBVS.5992....1D}[8]\\
2455978.87028 & 0.00070 & $V$ & 2021.5 & -0.0025 & \cite{2012IBVS.6029....1D}[9]\\
2456038.68868 & 0.00070 & $V$ & 2197.5 & -0.0024 & \cite{2012IBVS.6029....1D}[9]\\
2457811.65388 & 0.00010	& CCD-Clear & 7414 & -0.0045 & \cite{2021OEJV..211....1L}[10]\\
2457840.37499 & 0.00020	& CCD-Clear & 7498.5 & -0.0030 & \cite{2021OEJV..211....1L}[10]\\
2457840.54279 & 0.00020	& CCD-Clear & 7499 & -0.0051 & \cite{2021OEJV..211....1L}[10]\\
2458166.48289 & 0.00010	& CCD-Clear & 8458 & -0.0069 & \cite{2021OEJV..211....1L}[10]\\
2458932.39593 & 0.00047	& $V$ & 10711.5 & -0.0062	& This Study\\
2458932.56657 & 0.00024	& $V$ & 10712	& -0.0055 & This Study\\
\hline
\hline
\end{tabular}
\end{center}
\label{tab1}
\end{table*}

\vspace{1.5cm}
\section{Light Curve Solution}
The PHOEBE 2.4.9 version and the MCMC method were used to analyze the light curve of the V0610 Vir system (\citealt{2020ApJS..250...34C}[11]).

We used the $P-T_1$ relationship from the \cite{2022MNRAS.510.5315P}[12] study to calculate the effective temperature of the hotter star as the input (Equation \ref{eq2}).

\begin{equation}\label{eq2}
T_1=(6951.42 _{\rm-112.68}^{+112.16})P+(3426.01 _{\rm-43.90}^{+44.12})
\end{equation}

The gravity-darkening coefficients was determined $g_1=g_2=0.32$ (\citealt{1967ZA.....65...89L}[13]) and the bolometric albedo was assumed to be  $A_1=A_2=0.5$ (\citealt{1969AcA....19..245R}[14]). Additionally, the stellar atmosphere was modeled using the \cite{2004A&A...419..725C}[15] method, and the limb darkening coefficients were employed in the PHOEBE as a free parameter.

Due to the availability of photometric data, we used $q$-search to estimate the mass ratio. The obtained mass ratio was used as the MCMC process's initial parameter value.

The maxima of the light curve were asymmetric $(V_{max}1-V_{max}2)\neq0$. So, the light curve solution required the use of a hot starspot on the hotter component (\citealt{1951PRCO....2...85O}[16]).

\begin{figure*}
\begin{center}
\includegraphics[width=\textwidth]{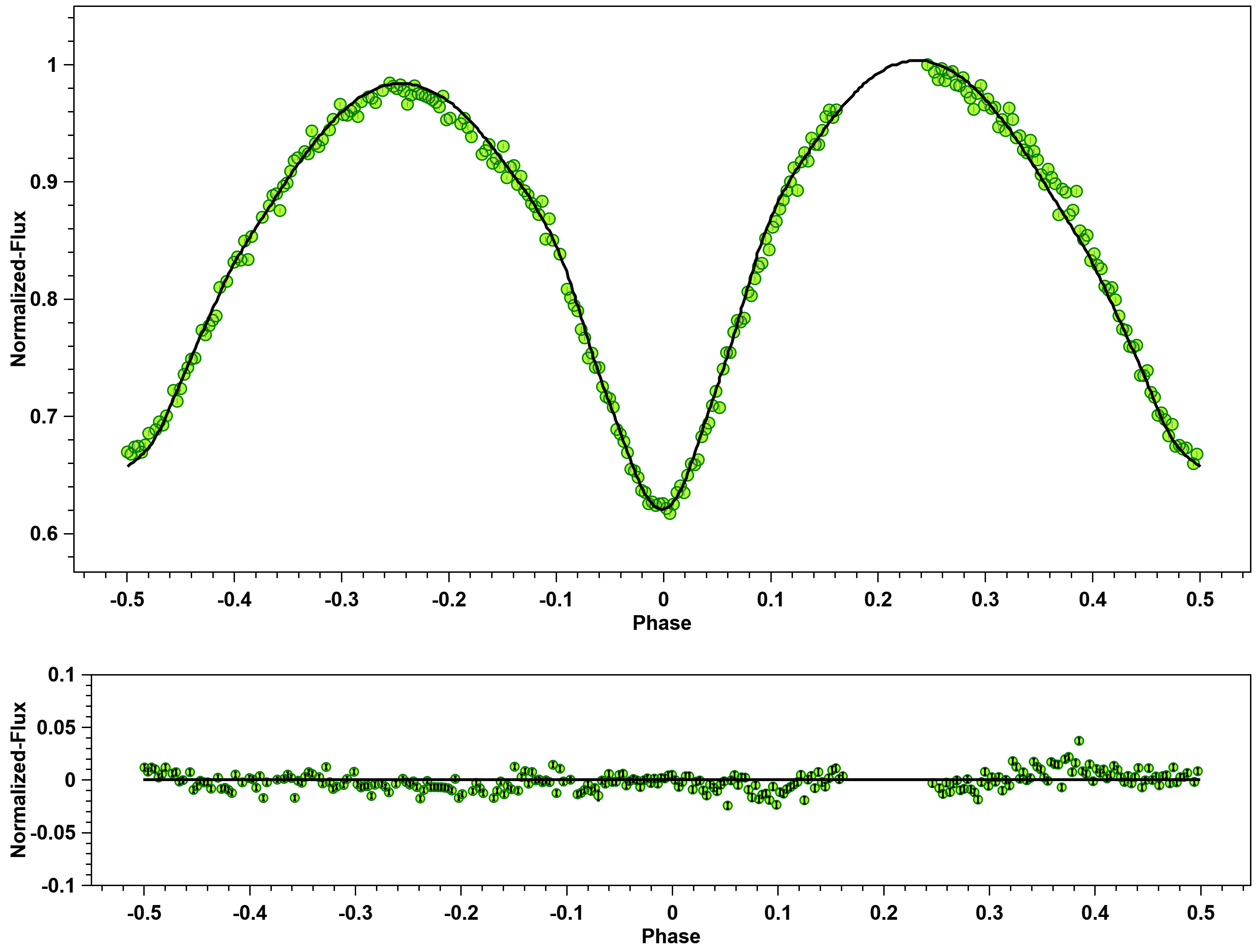}
    \caption{The observed and synthetic light curves of the system in $V$ filter.}
\label{Fig3}
\end{center}
\end{figure*}

\begin{figure*}
\begin{center}
\includegraphics[scale=0.65]{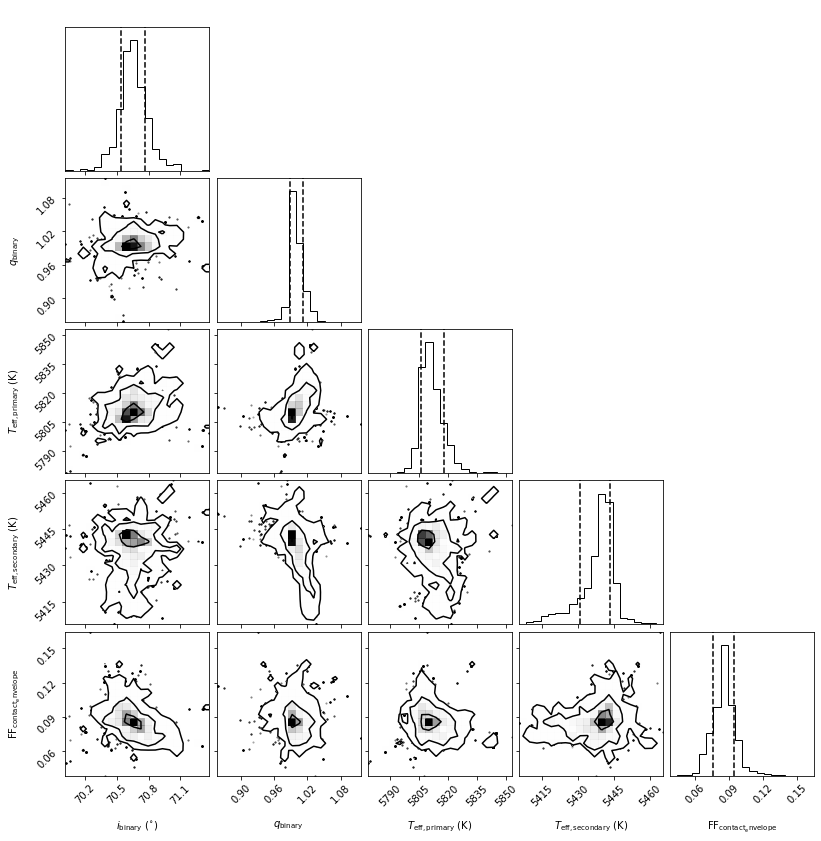}
    \caption{The corner plots of the system from the MCMC modeling.}
\label{Fig4}
\end{center}
\end{figure*}

\begin{figure*}
\begin{center}
\includegraphics[scale=0.54]{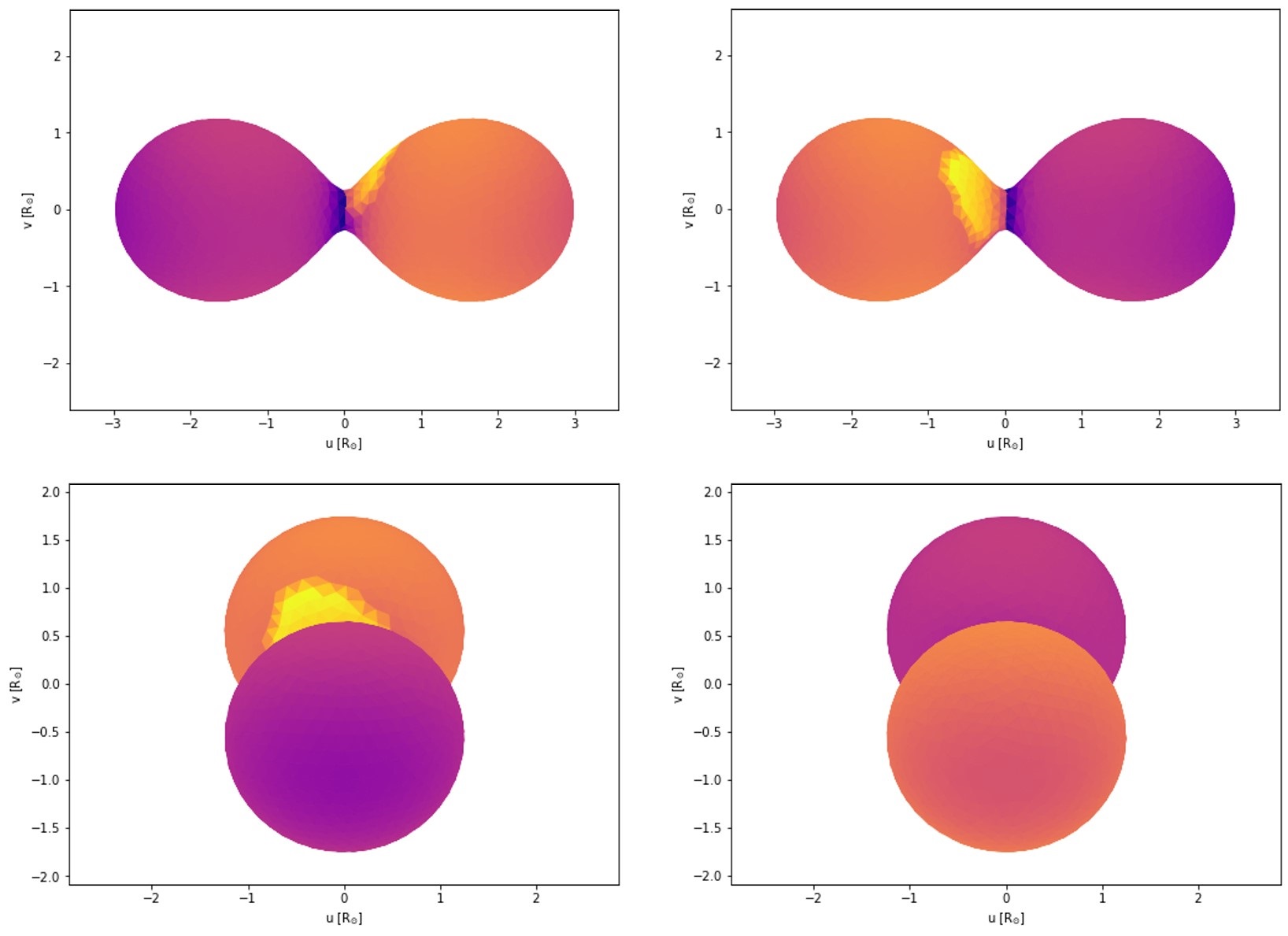}
    \caption{Geometric structure of V0610 Vir with a hot spot on the hotter component.}
\label{Fig5}
\end{center}
\end{figure*}

Then, the theoretical fit was improved using PHOEBE's optimization tool. Moreover, taking into account a normal Gaussian distribution in the range of solutions for inclination, mass ratio, fillout factor, and effective temperatures, we estimated the values of the parameters together with their uncertainties using the MCMC approach based on the emcee package in PHOEBE code (\citealt{2018ApJS..236...11H}[17]). We employed 96 walkers and 600 iterations for each walker in the MCMC processing.
Table \ref{tab2} contains the results of the light curve solution. The corner plots and final synthetic light curve are shown in Figure \ref{Fig3} and Figure \ref{Fig4}, respectively. The component positions for the four phases of an orbital period are shown in Figure \ref{Fig5}.

\begin{table}
\caption{Photometric solution of V0610 Vir.}
\centering
\begin{center}
\footnotesize
\begin{tabular}{c c c}
 \hline
 \hline
Parameter && Result\\
\hline
$T_{1}$ (K) && $5811_{\rm-(5)}^{+(7)}$\\
\\
$T_{2}$ (K) && $5440_{\rm-(9)}^{+(4)}$\\
\\
$q=M_2/M_1$ && $0.998_{\rm-(9)}^{+(15)}$\\
\\
$\Omega_1=\Omega_2$ && $3.70(5)$\\
\\
$i^{\circ}$ &&	$70.65_{\rm-(11)}^{+(12)}$\\
\\
$f$ && $0.085_{\rm-(9)}^{+(9)}$\\
\\
$l_1/l_{tot}(V)$ && $0.580(2)$\\
$l_2/l_{tot}(V)$ && $0.420(2)$\\
$r_{1(mean)}$ && $0.388(27)$\\
$r_{2(mean)}$ && $0.388(26)$\\
Phase shift && $-0.009(1)$\\
\hline
$Colatitude_{spot}(deg)$ && 74(1)\\
$Longitude_{spot}(deg)$ && 348(2)\\
$Radius_{spot}(deg)$ && 27(1)\\
$T_{spot}/T_{star}$ && 1.09(1)\\
$Radius_{spot}(deg)$ && 27(1)\\
Component && Hotter Star\\
\hline
\hline
\end{tabular}
\end{center}
\label{tab2}
\end{table}

\vspace{1.5cm}
\section{Absolute Parameters Estimation}
The estimation of absolute parameters in contact binary stars typically involves determining quantities such as mass, radius, luminosity, absolute bolometric magnitude, and surface gravity. So, to estimate absolute parameters, the Gaia DR3 parallax method was used in this study (\citealt{2021AJ....162...13L}[18], \citealt{2022MNRAS.510.5315P}[12]).
Therefore, we used $V_{max}=13.33\pm0.09$ from our observations, the extinction coefficient $A_v=0.063\pm0.001$ from the \cite{2011ApJ...737..103S}[19] study, and the Gaia DR3 distance $d(pc)=479\pm3$ for V0610 Vir.

So, we estimated the absolute magnitude $M_v$ of the system, subsequently the absolute magnitude for the components, and the bolometric magnitude $M_{bol}$ of each star, respectively (Equations \ref{eq3} to \ref{eq5}).

\begin{equation}\label{eq3}
M_v=V-5log(d)+5-A_v
\end{equation}

\begin{equation}\label{eq4}
M_{v(1,2)}-M_{v(system)}=-2.5log(\frac{l_{(1,2)}}{l_{(tot)}})
\end{equation}

\begin{equation}\label{eq5}
M_{bol}=M_{v}+BC
\end{equation}

The hotter and cooler components' bolometric correction $BC_1=-0.073$ and $BC_2=-0.153$ were derived as a function of the effective temperature of the stars (\citealt{1996ApJ...469..355F}[20]).

The following equations were used to determine the luminosity and radius, and separation between the center of mass of the components (\ref{eq6}, \ref{eq7} and \ref{eq8}):

\begin{equation}\label{eq6}
M_{bol}-M_{bol_{\odot}}=-2.5log(\frac{L}{L_{\odot}})
\end{equation}

\begin{equation}\label{eq7}
R=(\frac{L}{4\pi \sigma T^{4}})^{1/2}
\end{equation}

\begin{equation}\label{eq8}
a=\frac{R}{r_{mean}}
\end{equation}

Additionally, using the mass ratio determined by the outcomes of the light curve analysis, each component's mass was determined using the well-known Kepler's third law (Equation \ref{eq9}). Using Equation \ref{eq10}, the surface gravity was estimated.
The estimated parameters using the Gaia DR3 parallax are shown in Table \ref{tab3}.

\begin{equation}\label{eq9}
\frac{a^3}{G(M_1+M_2)}=\frac{P^2}{4\pi^2}
\end{equation}

\begin{equation}\label{eq10}
g=G_{\odot}(M/R^2)
\end{equation}

\begin{table}
\caption{The absolute parameters of the V0610 Vir binary system.}
\centering
\begin{center}
\footnotesize
\begin{tabular}{c c c c c}
 \hline
 \hline
Parameter & & Hotter star & & Cooler star\\
\hline
$M_v(mag.)$ && 5.457(72) && 5.807(70)\\
$M_{bol}(mag.)$ && 5.384(72) && 5.654(70)\\
$L(L_\odot)$ && 0.553(35) && 0.431(27)\\
$R(R_\odot)$ && 0.735(21) && 0.741(21)\\
$M(M_\odot)$ && 0.400(44) && 0.399(49)\\
$log(g)(cgs)$ && 4.307(20) && 4.300(26)\\
$a(R_\odot)$ && \multicolumn{3}{c}{1.902(71)}\\
\hline
\hline
\end{tabular}
\end{center}
\label{tab3}
\end{table}

\vspace{1.5cm}
\section{Summery and Conclusion}
We observed the V0610 Vir binary system at an observatory in the Czech Republic. We extracted our observed minima in addition to collecting from the literature. Then, we determined the epoch and O-C values using the reference ephemeris. The O-C diagram shows that just a liner fit can be considered, and that is descending.

Light curve analysis was performed using the latest available version of PHOEBE Python code together with the MCMC approach. Moreover, the Gaia DR3 parallax was used to estimate the absolute parameters of the V0610 Vir system.

According to the light curve analysis, the companion stars in this system have a temperature difference of 371 K. In contact systems, the maximum temperature difference between two stars is around 5\%, which is consistent with our light curve analysis in this regard (\citealt{2021RAA....21..203P}[21]).
Based on the temperatures of the stars, G3 and G8 are the spectral types of the hotter and cooler stars in this system, respectively (\citealt{2018MNRAS.479.5491E}[22]).

\begin{figure*}
\begin{center}
\includegraphics[scale=0.46]{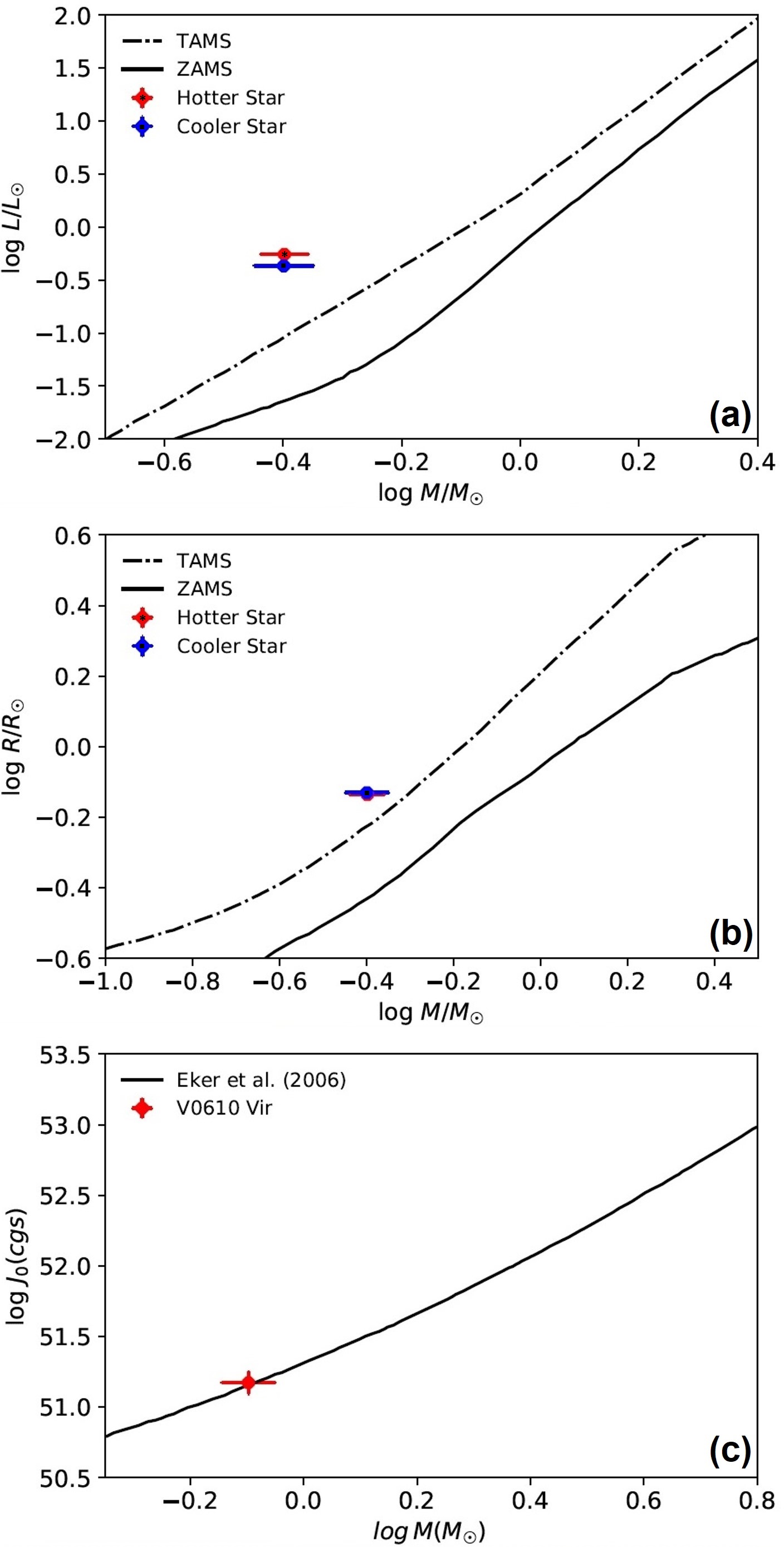}
    \caption{a) $logM-logL$ diagram; b) $logM-logR$ diagram; c) $logM_{tot}-J_0$ diagram.}
\label{Fig6}
\end{center}
\end{figure*}

The evolution of V0610 Vir is depicted by the positions of each component on the logarithmic scaled Mass-Luminosity ($M-L$) and Mass-Radius ($M-R$) diagrams (Figure \ref{Fig6}a,b). These diagrams show both the Terminal-Age Main Sequence (TAMS) and the Zero-Age Main Sequence (ZAMS). Due to their very close masses and radii, their position is next to each other and above TAMS.

The orbital angular momentum of the system is $51.173\pm0.079$. This result is based on the following equation from the \cite{2006MNRAS.373.1483E}[23] study:

\begin{equation}\label{eq11}
J_0=\frac{q}{(1+q)^2} \sqrt[3] {\frac{G^2}{2\pi}M^5P}
\end{equation}

where $q$ is the mass ratio ($M2/M1$), $M$ is the total mass of the system ($M1+M2$), $P$ is the orbital period, and $G$ is the gravitational constant. The units in equation \ref{eq11} are based on CGS.
The $logM_{tot}-J_0$ diagram (Figure \ref{Fig6}c) considers the V0610 Vir in a contact binary systems region.
\\
\\
According to the short orbital period, light curve solution, and estimation of the absolute parameters of the V0610 Vir, it can be concluded that this system is a Low-Mass Contact Binary (LMCB) system. The orbital period variation trend in LMCB binary systems is usually decreasing, so the examination of this requires more observations and a parabola on the O-C diagram. It should be noted that the LMCB systems have formed a disc that has the potential to be a place for planet formation with an age considerably shorter than the age of host stars (\citealt{2012AcA....62..153S}[24]).
So, based on the position of this system's stars in Figure \ref{Fig6} diagrams and the low mass of the two stars, we suggest it for future observations and investigations.

\vspace{1.5cm}
\section*{Acknowledgements}
This manuscript was prepared by the BSN project (\url{https://bsnp.info/}). We have made use of Gaia DR3 results. The Gaia mission is from the European Space Agency (ESA) (\url{http://cosmos.esa.int/gaia}), processed by the Gaia Data Processing and Analysis Consortium (DPAC).



\vspace{1.5cm}

\begin{thebibliography}{}
\bibitem[\protect\citeauthoryear{Kjurkchieva et al.}{2016}]{2016RAA....16..135K} [1] Kjurkchieva D.~P., Popov V.~A., Vasileva D.~L., Petrov N.~I., 2016, RAA, 16, 135. doi:10.1088/1674-4527/16/9/135
\bibitem[\protect\citeauthoryear{Zhang, Han, \& Liu}{2016}]{2016PASP..128l4201Z} [2] Zhang Y., Han Q.~W., Liu J.~Z., 2016, PASP, 128, 124201. doi:10.1088/1538-3873/128/970/124201
\bibitem[\protect\citeauthoryear{Poro et al.}{2024}]{2024PASP..136b4201P} [3] Poro A., Tanriver M., Michel R., Paki E., 2024, PASP, 136, 024201. doi:10.1088/1538-3873/ad1ed3
\bibitem[\protect\citeauthoryear{Li et al.}{2018}]{2018PASP..130g4201L} [4] Li K., Xia Q.-Q., Hu S.-M., Guo D.-F., Chen X., 2018, PASP, 130, 074201. doi:10.1088/1538-3873/aac067
\bibitem[\protect\citeauthoryear{Kazarovets et al.}{2015}]{2015IBVS.6151....1K} [5] Kazarovets E.~V., Samus N.~N., Durlevich O.~V., Kireeva N.~N., Pastukhova E.~N., 2015, IBVS, 6151, 1
\bibitem[\protect\citeauthoryear{Collins et al.}{2017}]{2017AJ....153...77C} [6] Collins K.~A., Kielkopf J.~F., Stassun K.~G., Hessman F.~V., 2017, AJ, 153, 77. doi:10.3847/1538-3881/153/2/77
\bibitem[\protect\citeauthoryear{Diethelm}{2010}]{2010IBVS.5945....1D} [7] Diethelm R., 2010, IBVS, 5945, 1
\bibitem[\protect\citeauthoryear{Diethelm}{2011}]{2011IBVS.5992....1D} [8] Diethelm R., 2011, IBVS, 5992, 1
\bibitem[\protect\citeauthoryear{Diethelm}{2012}]{2012IBVS.6029....1D} [9] Diethelm R., 2012, IBVS, 6029, 1
\bibitem[\protect\citeauthoryear{Lehk{\'y} et al.}{2021}]{2021OEJV..211....1L} [10] Lehk{\'y} M., {\v{S}}melcer L., Souza de Joode M., J{\'\i}lek F., et al., 2021, OEJV, 211, 1. doi:10.5817/OEJV2021-0211
\bibitem[\protect\citeauthoryear{Conroy et al.}{2020}]{2020ApJS..250...34C} [11] Conroy K.~E., Kochoska A., Hey D., Pablo H., et al., 2020, ApJS, 250, 34. doi:10.3847/1538-4365/abb4e2
\bibitem[\protect\citeauthoryear{Poro et al.}{2022a}]{2022MNRAS.510.5315P} [12] Poro A., Sarabi S., Zamanpour S., et al., 2022, MNRAS, 510, 5315. doi:10.1093/mnras/stab3775
\bibitem[\protect\citeauthoryear{Lucy}{1967}]{1967ZA.....65...89L} [13] Lucy L.~B., 1967, ZA, 65, 89
\bibitem[\protect\citeauthoryear{Ruci{\'n}ski}{1969}]{1969AcA....19..245R} [14] Ruci{\'n}ski S.~M., 1969, AcA, 19, 245
\bibitem[\protect\citeauthoryear{Castelli \& Kurucz}{2004}]{2004A&A...419..725C} [15] Castelli F., Kurucz R.~L., 2004, A\&A, 419, 725. doi:10.1051/0004-6361:20040079
\bibitem[\protect\citeauthoryear{O'Connell}{1951}]{1951PRCO....2...85O} [16] O'Connell D.~J.~K., 1951, PRCO, 2, 85
\bibitem[\protect\citeauthoryear{Hogg \& Foreman-Mackey}{2018}]{2018ApJS..236...11H} [17] Hogg D.~W., Foreman-Mackey D., 2018, ApJS, 236, 11. doi:10.3847/1538-4365/aab76e
\bibitem[\protect\citeauthoryear{Li et al.}{2021}]{2021AJ....162...13L} [18] Li K., Xia Q.-Q., Kim C.-H., et al., 2021, AJ, 162, 13. doi:10.3847/1538-3881/abfc53
\bibitem[\protect\citeauthoryear{Schlafly \& Finkbeiner}{2011}]{2011ApJ...737..103S} [19] Schlafly E.~F., Finkbeiner D.~P., 2011, ApJ, 737, 103. doi:10.1088/0004-637X/737/2/103
\bibitem[\protect\citeauthoryear{Flower}{1996}]{1996ApJ...469..355F} [20] Flower P.~J., 1996, ApJ, 469, 355. doi:10.1086/177785
\bibitem[\protect\citeauthoryear{Poro et al.}{2021}]{2021RAA....21..203P} [21] Poro A., Alicavus F., Fern{\'a}ndez-Laj{\'u}s E., et al., 2021, RAA, 21, 203. doi:10.1088/1674-4527/21/8/203
\bibitem[\protect\citeauthoryear{Eker et al.}{2018}]{2018MNRAS.479.5491E} [22] Eker Z., Bak{\i}{\c{s}} V., Bilir S., Soydugan F., et al., 2018, MNRAS, 479, 5491. doi:10.1093/mnras/sty1834
\bibitem[\protect\citeauthoryear{Eker et al.}{2006}]{2006MNRAS.373.1483E} [23] Eker Z., Demircan O., Bilir S., Karata{\c{s}} Y., 2006, MNRAS, 373, 1483. doi:10.1111/j.1365-2966.2006.11073.x
\bibitem[\protect\citeauthoryear{St{k{e}}pie{\'n} \& Gazeas}{2012}]{2012AcA....62..153S} [24] St{k{e}}pie{\'n} K., Gazeas K., 2012, AcA, 62, 153. doi:10.48550/arXiv.1207.3929
\end{thebibliography}
\end{document}